# Engineering magnetic domain-wall structure in Permalloy nanowires

M. J. Benitez[1,*], M. A. Basith[1,†], D. McGrouther[1], S. McFadzean[1], D. A. MacLaren[1], A. Hrabec[2], R. J. Lamb[1], C. H. Marrows[2] and S. McVitie[1]

[1]SUPA, School of Physics and Astronomy, University of Glasgow, Glasgow G12 8QQ, United Kingdom

[2]School of Physics and Astronomy, University of Leeds, Leeds LS2 9JT, United Kingdom

---

[*] Email: maria.benitez-romero@glasgow.ac.uk

[†] Present address: Department of Physics, Bangladesh University of Engineering and Technology, Dhaka-1000, Bangladesh

Using a focused ion beam microscope we have created non-topographic features that provide controlled modification of domain wall structure, size and pinning strength in 500nm wide nanowires composed from Cr(3 nm)/Permalloy(10 nm)/Cr(5 nm). The pinning sites consist of linear defects where magnetic properties were modified by a $Ga^+$ ion probe of diameter ~ 10 nm. Detailed studies of the structural, chemical and magnetic changes induced by the irradiation, which showed the modified region to be ~40-50nm wide, were performed using scanning transmission electron microscopy modes of bright field imaging, electron energy loss spectroscopy and differential phase contrast imaging on an aberration (Cs) corrected instrument. The Fresnel mode of Lorentz transmission electron microscopy, was used for studies of domain wall behaviour, where we have observed changes in depinning strength and structure with irradiation dose and line orientation. We present an understanding of this behaviour based upon micromagnetic simulation of the irradiated defects and their effect on the energy terms for the DWs.

PACS numbers: 75.60.Ch, 75.50.Bb, 68.37.Lp, 62.23.Hj

## I. INTRODUCTION

The controlled manipulation of magnetic domain walls (DWs) in laterally constrained ferromagnets is of huge interest for applications in logic devices [1], spin oscillators [2] and data storage [3]. The domain wall structure in soft magnetic materials is determined by the width and thickness of the nanowire [4,5] (usually < 20 nm for the thickness and < 500 nm for the width). The most commonly investigated soft material, permalloy ($Ni_{80}Fe_{20}$), has well-studied domain wall structures, the transverse wall (TW) and vortex wall (VW), which are extended objects, having dimensions comparable with the width of the wire [4]. Considerable attention has been devoted to fabricate geometric modifications in those nanowires such as notches/anti-notches of different geometries to control not only the DW structure, but also the DW pinning strength at the modified sites [6-9].

Using localized $Ga^+$ ion irradiation we have previously demonstrated that DWs could be reproducibly pinned at non-topographic sites created by irradiation [10] of a tri-layer nanowire of Cr/Permalloy(Py)/Cr and that the pinning strength strongly depended on the irradiation dose used. In this article, we extend this investigation, demonstrating that it is possible to select the type, chirality and size of the DW in Py nanowires at the non-topographic sites created by Ga$^+$ ion irradiation. A Ga$^+$ ion probe with full width at half-maximum (FWHM) diameter ~ 10 nm in the focused ion beam (FIB) microscope is used to intermix the layers such that Cr becomes distributed within the Py layer [11]. Small additions of Cr are known to significantly reduce the saturation magnetization and Curie temperature of ferromagnetic NiFe alloys, which become paramagnetic at room temperature with only about 8 at% Cr [11, 12]. Furthermore, we show that the localized changes of the order 2-3× diameter of the focused ion spot can be measured using the differential phase contrast (DPC) imaging technique on an aberration Cs corrected scanning transmission electron microscope (STEM). This provides the capability of quantitative measurements with nanometre spatial resolution [13]. Using the Fresnel mode of Lorentz microscopy and micromagnetic simulations we study the local changes in domain wall behavior for four different line orientations, i.e. 30, 45, 67.5 and 112.5 degrees with respect to the wire length and five irradiation doses of $d \times 10^{15}$ ions/cm$^2$ ($d$ =4, 8, 12, 16 and 20). The irradiation dose was chosen taking into account the results from the dynamic ion irradiation simulation package TRIDYN [14]. These simulations showed that 8 at% Cr alloyed into the NiFe layer is achieved by an ion dose of $16 \times 10^{15}$ ions/cm$^2$ (see supplementary information). Thus, ferromagnetism is expected to vanish and the material becomes paramagnetic at this dose.

This article is organized as follows. In Section II, we describe the experimental details of the material and techniques used as well as the parameters of micromagnetic simulations. In Section III, we discuss the results obtained in this work. We start by exploring the structural and magnetic modification of large areas, with width 1 μm and length 10 μm, of thin film caused by Ga$^+$ ion irradiation. Next, we concentrate on the effect of the irradiation on the structural and magnetic properties for a single line written with a Ga$^+$ probe of 10 nm diameter. This local modification is proven to be an efficient method to stabilize DW structures that are quite different to the equilibrium structure that one would find free in the nanowire and to control its chirality and pinning strength. Finally, Section IV contains the summary and the conclusions of the study.

## II. EXPERIMENTAL DETAILS

A thin film stack of Cr(3 nm)/Py(10 nm)/Cr(5 nm) was grown using DC sputtering in Ar plasma on top of an electron transparent $Si_3N_4$ window substrate suitable for transmission electron microscopy (TEM) studies [15]. Nanowires were fabricated from the continuous films by milling the film stack using a FEI Nova NanoLab 200 scanning electron microscope/focused ion beam workstation using a 30 keV $Ga^+$ beam energy. The nanowires of 15 µm length and 500 nm width connected to a right-angled bend are shown schematically in Fig. 1(a). Linear defects were created using the FIB with an ion beam current of 10 pA and an ion probe size of about 10 nm in diameter. Nominally the ion irradiation doses ($d$) chosen were $d \times 10^{15}$ ions/cm$^2$ ($d$ =4, 8, 12, 16 and 20), which should cause a range of intermixing of the Cr and Py layers. This localized irradiated single line was oriented at angles of 30, 45, 67.5 or 112.5 degrees with respect to the wire length and was defined using the different $Ga^+$ doses. Fig. 1(b) shows a plan view TEM image of the nanowire with the irradiated line (highlighted in dotted blue) written at an angle of 67.5 degrees. Note that the line is written between the two heavily irradiated marker spots of diameter 300 nm that are highlighted in the figure, which help to identify the line's location.

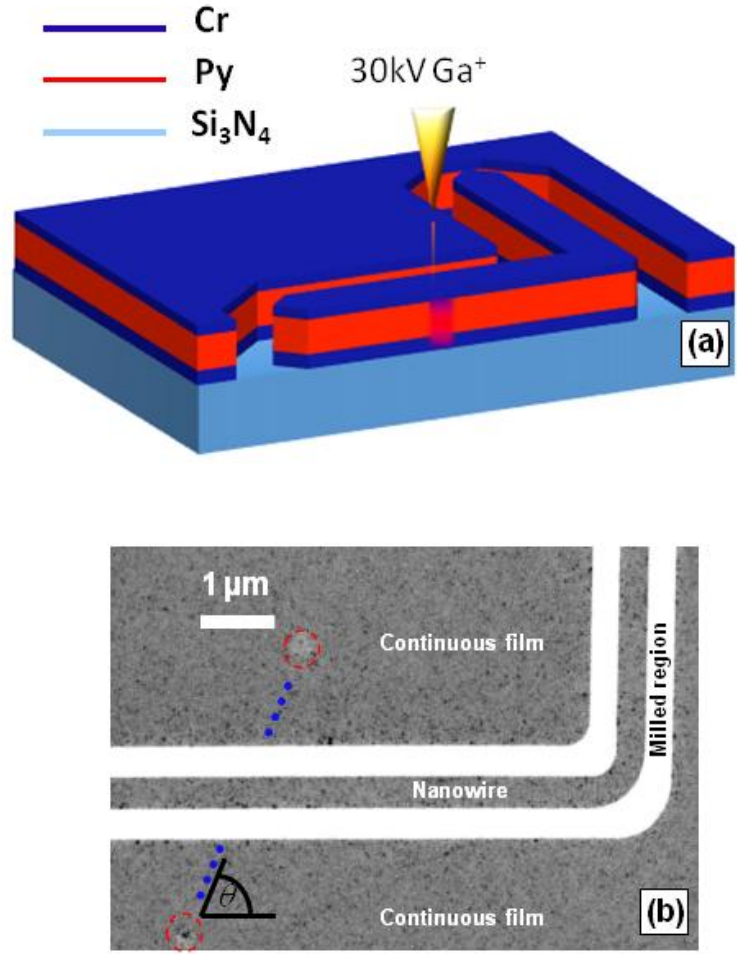

FIG 1. (color online). (a) Schematic of the fabrication process. (b) Plan view TEM image of the nanowire fabricated by FIB milling. The blue dotted line shows the orientation of the irradiated line which was written continuously between the two red open circles. θ is the angle between the irradiated line and the wire length.

The modification of the magnetic properties of the stack caused by the irradiation was investigated on a continuous film by creating an irradiated stripe pattern 1 μm wide and 10 μm in length employing the dose values noted earlier. The stripe patterns were created systematically across the continuous film on top of the $Si_3N_4$ window membrane and analyzed using bright field and diffraction imaging as shown in Section III.

Next, cross-sectional TEM samples containing irradiated lines of the different doses were prepared using a FIB-based 'in situ' lift-out technique [16,17] to observe the details of the physical structure and chemical compositional of the irradiated areas using electron-energy-loss spectroscopy (EELS). A comparison of the composition of irradiated versus non irradiated areas allowed a confirmation of the composition change induced by the Ga$^+$ beam that could be correlated with magnetic measurements. The effect of the magnetically modified line on the domain wall pinning was investigated using the DPC and Fresnel modes of Lorentz microscopy [18]. Fresnel images and low angle electron diffraction (LAD) patterns were obtained using a Philips CM20 microscope equipped with a field emission gun and designed for in-situ magnetization experiments, whereas DPC and EELS measurements were obtained using a JEOL ARM-200FCS aberration-corrected (Cs) scanning microscope, both operated at 200kV. EELS data were collected using a Gatan Quantum 965 spectrometer.

To investigate the energetics of domain wall structures and assist the interpretation of experimental observations obtained by Lorentz microscopy, micromagnetic simulations were carried out using the object oriented micromagnetic framework (OOMMF) code [19]. 500 nm wide and 10 nm thick nanowires were simulated to contain a modified region of 50 nm width oriented at 30, 45, 60 and 120 degrees. The width of the modified region was chosen to take into account the experimental results from cross-sectional TEM imaging of the irradiated region. The simulation cell size was 5×5×5 nm$^3$ for all simulations. The parameters used here were the standard for Py, i.e. the saturation magnetization $M_s$= 8.6 × 10$^5$ A m$^{-1}$, exchange constant $A$=13 × 10$^{-12}$ J m$^{-1}$ and zero magnetic anisotropy constant. The saturation magnetization of the modified region ($M_{si}$) was varied from 90% to 0% of the standard Py value. The damping coefficient α was set to 0.5 in order to quickly converge the static magnetic structure.

## III. RESULTS AND DISCUSSION

### A. Effects of the irradiation on the thin film

Figure 2 shows plan view TEM bright field images of (a) the as-deposited unirradiated continuous multi-layer film and (b) a region of alternating stripes of unirradiated and (1 µm wide) irradiated material. The irradiated region can be distinguished from the unirradiated film by the increase in grain size, from 5-10 nm in the unirradiated film to 20-30 nm in the irradiated region, which is consistent with previous reported studies [20,21].

Quantitative measurements of the magnetic changes due to irradiation in the stripes were obtained using LAD by direct observation of the magnetic structure in the diffraction pattern. LAD manifests as a small angle splitting of the central diffraction spot due to the change in magnetic induction across a DW in the sample. It may be interpreted classically as a spectrum of the Lorentz deflection angles and magnitudes present in the region of specimen under investigation in TEM. The Lorentz deflection angle β of electrons in magnetic materials is given by $\beta_L = \frac{e\lambda B_s t}{h}$, where $e$ is the electronic charge, $h$ is Planck's constant, $t$ is the film thickness, $B_s$ is the saturation induction and $\lambda$ is the electron wavelength [18,22]. This calculation assumes that the induction/magnetization lies in the plane of the film, perpendicular to the incident electron beam. For 20 nm thick Py film ($B_s$ = 1 T) and 200 kV accelerating voltage (i.e. $\lambda$ = 2.5 pm) β is 12.7 µrad, which is several orders of magnitude smaller than the smallest Bragg diffraction angle of the crystalline lattice.

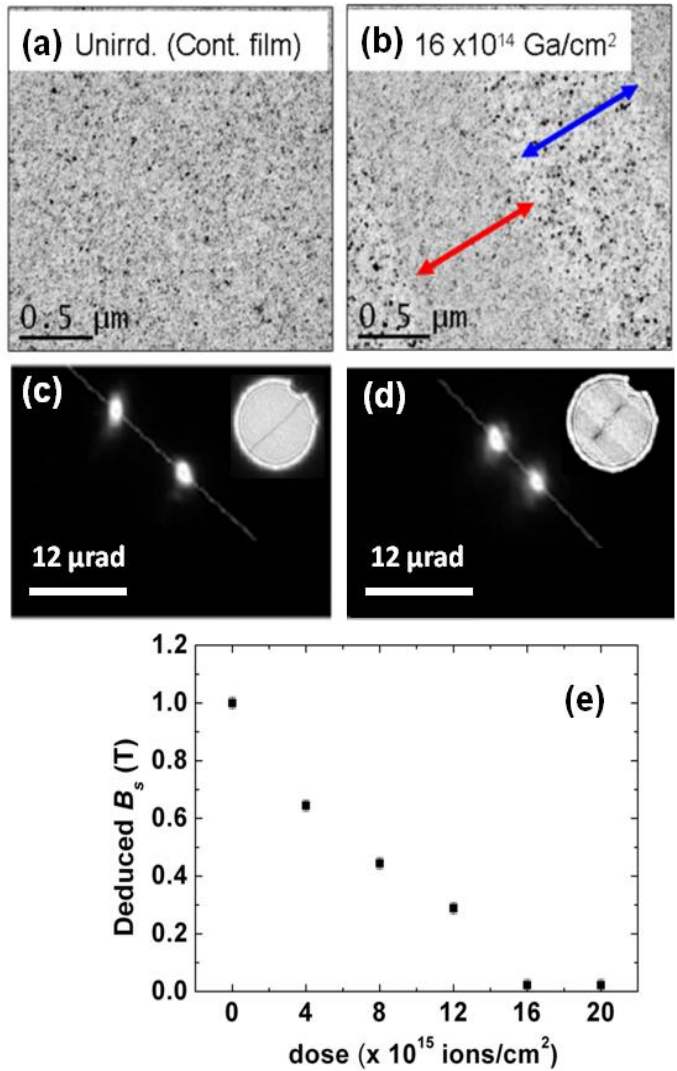

FIG 2. (color online). TEM bright field images of the (a) unirradiated continuous film and (b) alternative unirradiated/irradiated stripe patterns, indicated by red and blue arrows respectively. Diffraction patterns from (c) unirradiated continuous film and (d) irradiated regions (16 x $10^{15}$ ions/cm$^2$). Insets (c and d) show Fresnel image of a 180$^o$ domain wall in a continuous thin film. (e) Deduced $B_s$ as a function of the irradiation ion doses.

Low angle diffraction patterns taken from the films, where a 180$^o$ domain wall was nucleated, were recorded for the unirradiated region (Fig. 2(c)) and from regions modified by different irradiation doses. Figure 2(d) shows the diffraction pattern for an irradiation dose of $16\times10^{15}$ ions/cm$^2$. One observes that the separation between the two diffraction spots is smaller for the deflection which takes place from a region containing irradiated stripes compared to that of the unirradiated region, confirming the reduction of magnetization. By measuring the separation of the spots and using the equation, $\beta_L = \frac{e\lambda B_S t}{h}$, the value of $B_s$ can be quantitatively determined and thus the effect of the irradiation deduced. Figure 2(e) shows how the deduced $B_s$ decreases as a function of ion dose for the irradiated stripe pattern. As can be seen from this graph the values range from 1 T for the unirradiated film to 0 T for a dose of $16\times10^{15}$ ions/cm$^2$, indicating that the materials has been rendered non ferromagnetic at this dose in agreement with TRIDYN simulations.

### B. Effects of the localized irradiation

The effect of the irradiation of a single line written with a probe of 10 nm diameter was first investigated using a cross-sectional TEM sample. Figure 3(a) shows a dark field STEM image of the cross-section, with contrast deriving principally from atomic number so that the protective Pt capping layer deposited in the FIB system appears brightest. There is insufficient contrast to distinguish Py from Cr, although the layers are readily observed by EELS, as described below. A small depression in the upper Cr layer (indicated by the dashed green arrow) indicates the irradiated region, here created with an ion dose of $16\times10^{15}$ ions/cm$^2$. The depression is consistent with a little sputtering and reduction of film thickness in the area of irradiation, in agreement with previous work [20]. Note that the apparent width of the irradiated region is approximately 50 nm, which is clearly larger than the 10 nm ion probe diameter. This broadening can be understood by considering the means by which the incident ions undergo energy loss in the material. The dominant mode of energy loss for the 30keV Ga$^+$ ions occurs through causing displacements of atoms in the sample. In these interactions the displaced atoms can possess excess kinetic energy so that they travel in random directions eventually colliding with other atoms in the sample. Thus, a region of altered material is created whose width is larger than the beam diameter and whose overall dimensions depend on factors such as incident beam energy, material density and average atomic number. Irradiation-induced intermixing of the Cr and Py layers was assessed by EELS, as summarized in Fig. 3(b) and (c), which show the elemental distributions within unirradiated (b) and irradiated (c) regions. The data were collected using the spectrum imaging technique [23], where an EELS spectrum was acquired pixel-by-pixel along the direction indicated by the red arrows in Fig. 3(a). The (normalized) EELS spectral intensity of the Cr, Fe and Ni $L_{2,3}$ excitation edges are plotted, following removal of a power-law background and assuming standard scattering cross-sections. The intensity of the O K-edge was also collected to check for oxidation effects. Processing was performed using the Gatan Digital Micrograph software package. The Cr-Py-Cr stack is apparent in both figures (b) and (c), with negligible oxygen within the stack. None of the layers have a "top-hat" profile and the tails of the distributions overlap, suggesting a degree of intermixing of Cr and Py even in the unirradiated region. However, apparent overlap of the distributions will also occur due to broadening of the scattered electron beam and (more importantly) because the data derive from a projection of a three-dimensionally-rough interface onto a two dimensional plane, as described elsewhere [24]. We therefore concentrate on the comparison between the two linescans, which shows a clear difference in the Cr distribution. Most notably, Cr is evident throughout the stack in the irradiated region and the Cr component does not dip below 10% atomic fraction. This finding suggests that the irradiated region at this dose is paramagnetic at room temperature, consistent with bulk measurements of alloying of Py and Cr [11]. Therefore the EELS data confirms the LAD measurements in Fig. 2(e): that when the film is irradiated with a dose of $16\times10^{15}$ ions/cm$^2$ and above, it becomes paramagnetic and loses its spontaneous magnetization.

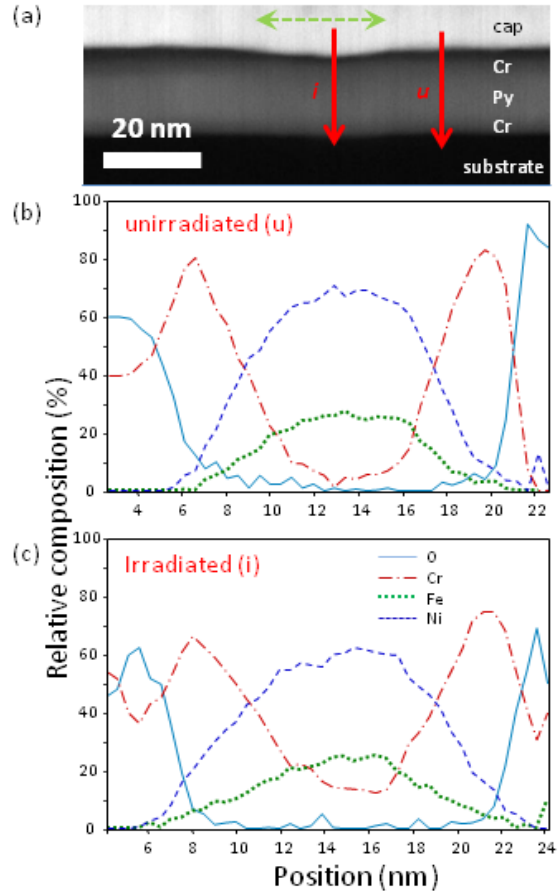

FIG 3. (color online). (a) High angle annular dark field (HAADF) STEM image of cross-section through an irradiated line in the Cr-Py-Cr stack. The ion dose in the irradiated region was $16 \times 10^{15}$ ions/cm$^2$ and a protective Pt capping layer was deposited prior to cross-section fabrication. The green arrows indicated the irradiated zone whilst the red arrows indicate the location of EELS linescans. (b,c) Elemental distributions derived from the EELS data and collected from unirradiated (u) and irradiated (i) regions, respectively.

Quantitative local measurements of the magnetic properties of the linear defects were explored by DPC imaging. In comparison to LAD, which provides global magnetic information from a relatively wide illuminated region, DPC imaging yields local integrated induction measurements at very high spatial resolution, below 10 nm [13,25]. DPC measurements were performed on the irradiated lines written between the nanowire edge and one heavily irradiated marked spot (Fig. 1b). The sample was uniformly magnetized by applying a magnetic field parallel to the long axis of the nanowire. Figures 4(a) and 4(b) show the experimental components of induction of an irradiated line at 45 degrees orientation written with a dose of $8\times10^{15}$ ions/cm$^2$. The DPC detector was aligned so that the components of induction mapped were orthogonal and parallel to the length of the irradiated line, as indicated by the double headed white arrows. One can clearly observe a contrast change in the irradiated line compared to outside this region in the component mapping along the line (Fig. 4(a)). This contrast reflects the local difference in magnetic induction. Mapping the component of induction perpendicular to the line, Fig. 4(b), shows no contrast variation. Such a result is entirely expected as the normal component of magnetic induction is continuous across a boundary in keeping with the requirement for $\nabla.\mathbf{B} = 0$. $\Delta\beta_L$ is measured by taking the difference of the mean value of $\beta_L$ of the unirradiated region and the irradiated one across an area of 400×100 nm$^2$ as indicated by the black dotted region in Fig. 4 (a). The profiles taken from the experimental and calculated images are shown in Fig. 4(c) whereas the $\Delta\beta_L$ measurements for all different irradiation doses are displayed in Fig. 4(d). The high standard deviation in $\Delta\beta_L$ in the experimental results arises due to diffraction effects from the individual crystallites of which the film is composed, which scatter electron intensity into Bragg diffraction channels and therefore affect the intensity of the undiffracted beam. This diffractive contrast has been reduced as much as possible by averaging over the area indicated in Fig. 4(a). We have assumed that the magnetization in the irradiated region is parallel to the length of the wire (the sample was saturated by applying a large field along the wire edge and then reducing this to zero) and comparing the experimental and calculated images we can conclude that for the irradiation doses of $16\times10^{15}$ ions/cm$^2$ and $20\times10^{15}$ ions/cm$^2$, there is enough Cr alloyed into the NiFe alloy film to change the ferromagnetic region to a paramagnetic one. This is consistent with the results obtained using LAD and EELS, however it shows that the local magnetic induction in the line has been quantitatively measured rather than a large area modified region which was deduced from the LAD measurements in Fig. 2. Furthermore the magnetization profile of the irradiated line is seen to have a width of around 50 nm, consistent with the HAADF cross-section image (Fig. 3(a)).

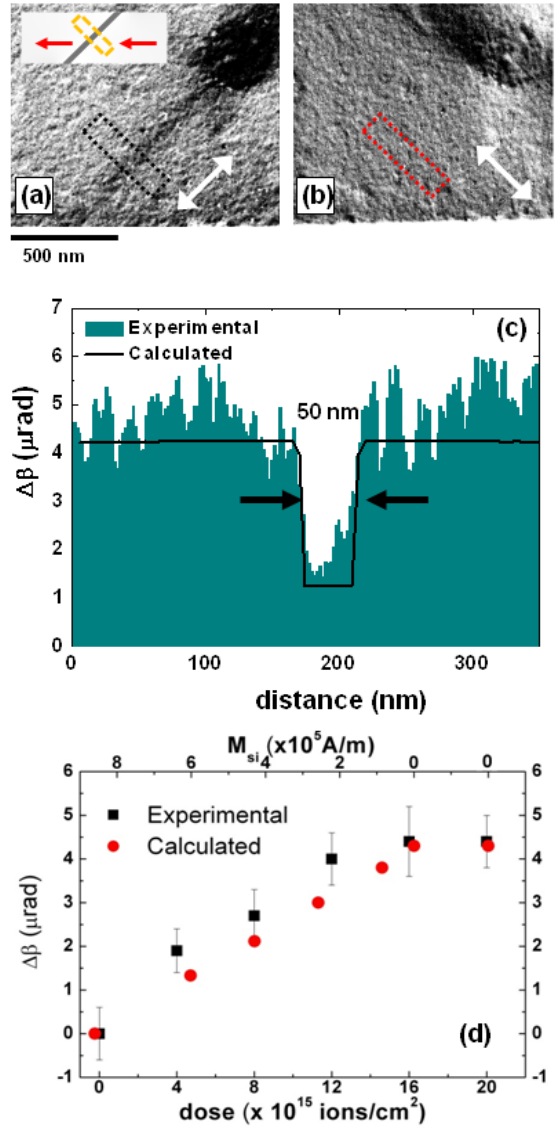

FIG 4. (color online). (a) and (b) Experimental DPC images from a linear pinning site of 45 degree orientation and irradiation dose of $8\times10^{15}$ ions/cm$^2$. The double headed white arrows indicate the direction of sensitivity of the mapped induction components. The inset shows calculated Lorentz DPC images from a linear pinning site of 45 degree orientation and $M_{si}=0.3M_s$. (c) Profiles taken from experimental (black dotted rectangle) and calculated images (yellow dotted rectangle) shown on (a). (d) Experimental and calculated $\Delta\beta_L$ vs the irradiation dose and $M_{si}$. Error bars correspond to the standard deviation of $\Delta\beta_L$.

### C. Observed Domain wall behavior

The magnetic behavior of the DWs pinned at linear defect regions with different orientations and saturation magnetization is now explored. First, a head-to-head TW was created at the bend in the structure as shown in Fig. 5(a) (with schematic shown in Fig. 5(b)). The initial TW was obtained by applying a magnetic field of around 600 Oe along the wire length and relaxing the field to zero. Next, a magnetic field of opposite sign was applied along the wire length to propagate the DW towards the modified region. The propagation field i.e., the field required to move the DW from the corner position to the pinning site was identical for all wires, being 8 ± 2 Oe as reported previously [10]. Figs. 5(c) and 5(d) show Fresnel images of the DW structure for linear pinning irradiated sites written with a dose of $4\times10^{15}$ ions/cm$^2$ and with two different orientations, i.e., 30 and 67.5 degrees. A head to head TW is observed in both cases. Note that the leading part of the TW is pinned along the line written at an angle of 67.5 degrees, whereas the wall is just pinned at one point of the pinning site oriented at 30 degrees. Fresnel images calculated from simulations where head-to-head DWs are artificially introduced in wires with local property modifications at two different orientations, 30 and 60 degrees and with $M_{si}=0.9M_s$ are included in Figs. 5(e) and 5(f). The red arrows in the calculated images depict the direction of the magnetization around the defect. The calculated Fresnel image, Fig. 5(e) also demonstrates that the leading part of the TW is pinned just where the line meets the lower edge of the wire as observed in the experiment, Fig. 5(c). For a line orientation of 60 degrees, calculated Fresnel image Fig. 5(f), the leading part of the wall is pinned along the defect as was observed in experimental Fresnel image, Fig. 5(d). By reducing the magnetization in the defect, i.e. for $M_{si} \leq 0.8M_s$ the DW is pinned along the modified region of the nanowire oriented at 30 degrees as seen in the simulated image, Fig. 5(g).

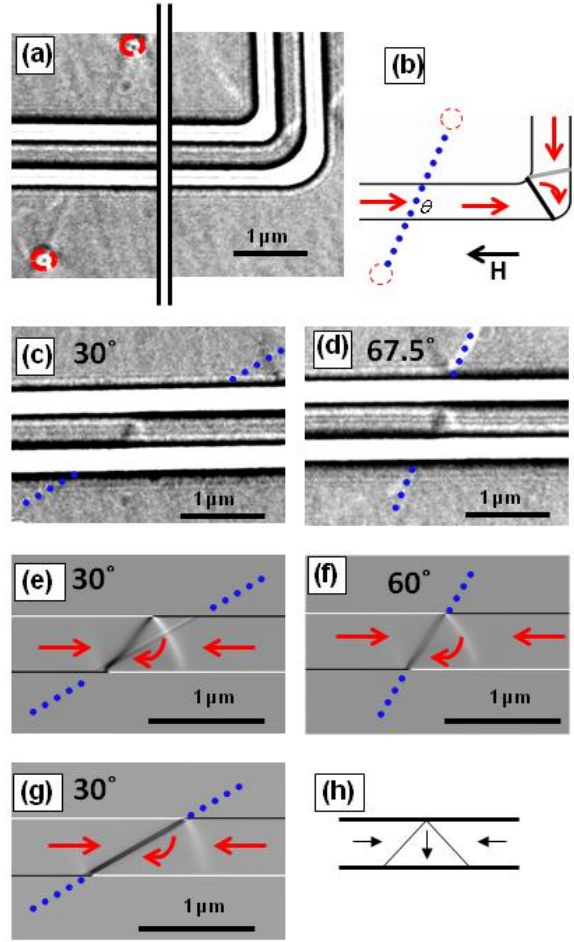

FIG 5. (color online). (a,c,d) Fresnel images of Cr/Py/Cr nanowires. (a) TW at zero field located at the bend in the nanowire and (b) corresponding schematics. (c and d) DWs pinned at the linear irradiated sites, written with a dose of $4\times10^{15}$ ions/cm$^2$ but with different orientations. (e-g) Calculated Fresnel images from simulations of the wires with local property modifications with $M_{si} = 0.9M_s$ (e, f) and $M_{si} = 0.8M_s$ (g). The blue dotted lines show the orientation of the irradiated lines whereas the red arrows represent the magnetization direction. Note the TW is pinned just where the line meets the lower edge of the wire for (c) and (e) whereas for (d, f, g), the leading part of the wall is pinned along the defect. (h) Schematic of a TW in a uniform part of a nanowire.

Experimentally a further reduction of magnetization at the modified region can be achieved by increasing the irradiation dose. Thus, irradiation doses of $d \times 10^{15}$ ions/cm$^2$ (with $d =$ 8, 12 and 16) were applied for all the linear orientations. Figures 6(a), 6(b) and 6(c) show Fresnel images of DWs pinned at the linear irradiated sites written with a dose of $16\times10^{15}$ ions/cm$^2$. As the irradiation dose increases there is a change of the DW structure and VW becomes the preferred wall type independent of the line orientation. The observed DW for higher doses with linear orientation smaller than 90 degrees is always a VW with a clockwise sense of circulation, or chirality. This is in agreement with previous observations [10] where only a 45$^o$ linear defect was investigated. In this case the central wall section of the VW aligns along the defect. For the cases observed here this wall section appears to show the wall extended (Fig. 6(a) 30$^o$) and compressed (Fig. 6(b) 67.5$^o$), which is consistent with the central wall section aligning with the defect in each case. Counterclockwise VWs are obtained for wires where the linear defect is written at an angle higher than 90 degrees as shown in Fig. 6(c). In this case, for the central wall section to align along the defect requires the opposite chirality VW compared to those at the lower angle. A simulation of this situation confirms the experimental findings (Figs. 6 (d-f)). Overall, this behavior is attributed to induced asymmetry in the energy landscape of the modified region of nanowire due to ion irradiation. Thus, the preferred chirality of the VW is that one for which the angle between the central wall section and the irradiated line is minimized.

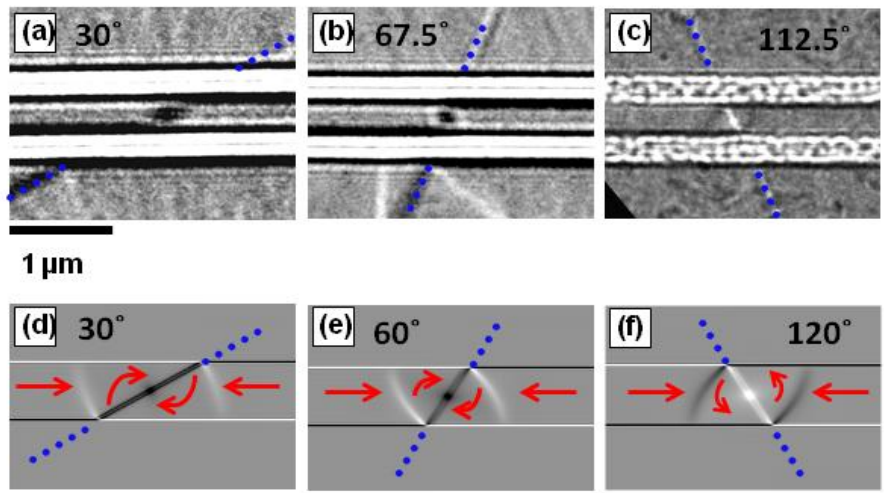

FIG 6. (color online). (a-c) Fresnel images of Cr/Py/Cr nanowires showing DWs pinned at the linear irradiated sites written with a dose of $16\times10^{15}$ ions/cm$^2$. (d-f) Corresponding calculated Fresnel images from simulations.

Furthermore, depinning field, $H_{dep}$, values were measured for different domain wall configurations. Table I ($H_{dep}$ direction is shown in schematics in table I) summarizes the results for an initial TW structure pointing down and lists the subsequent pinned wall structure and its depinning field as a function of the line dose and line orientation with the statistics included to show the variation in wall type observed at the defect.

### D. Discussion of energetics and of DW behaviour

Consideration of the energetics of domain wall structure and the influence by the irradiated defects is necessary to explain the observations we have made in Section C. For domain walls in straight nanowires, structure and energetics have been the subject of detailed micromagnetic study [4] that yielded a phase diagram predicting the ground state domain wall configuration based on wire geometry. Wider, thicker wires tended to favour vortex type walls, while thinner, narrower wires favour transverse walls. For the wire dimensions employed here, the expected ground energy state would be a vortex wall. In all of our experiments we have produced an initial transverse (down) domain wall through use of a polarizing field at the bend in the wire, Table I Fig. (a). The metastable transverse wall appears to remain stable in the propagating field although it has been seen to change its structure at the defect as we have observed both here and in our previous study [10].

Guided by the microscopy characterisation in Section B, we have conducted extensive micromagnetic simulations to understand the effects of the irradiated defects in pinning walls and transforming their structure. Figure 7 shows a plot of the total energy for pinned domain walls with respect to the magnetisation $M_{si}$ of the irradiated line defects with 30,45,60 degree angles. In agreement with Reference [4], at $M_{si}=8.6 \times 10^5$ A/m, vortex type walls are predicted to have a ~10% lower energy than transverse walls for this wire geometry. For both wall types, reducing the local strength of the magnetisation ($M_{si}$ decreasing from $8.6 \times 10^5$ A/m to zero) in the line defect leads to a monotonic reduction of the total DW energy (reducing both magnetostatic and exchange energy terms) with irradiation dose. This constitutes the creation of a potential energy well. Any displacement of the wall away from the irradiated defect immediately leads to an increase in the total energy.

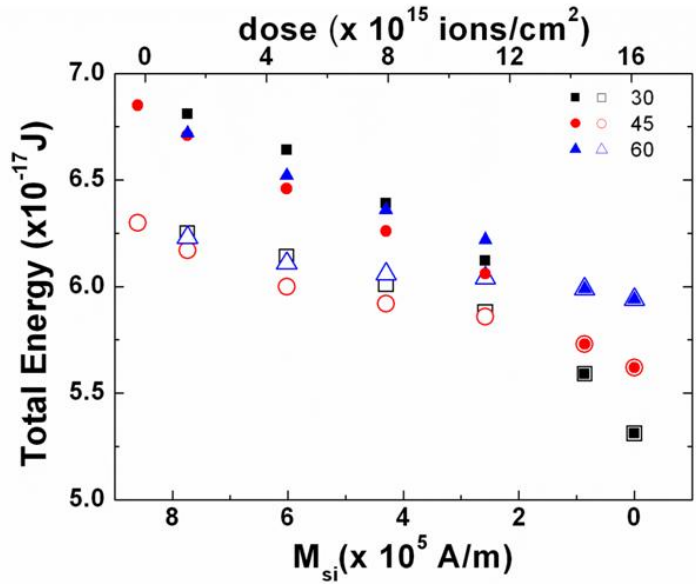

FIG 7. (color online). Calculated Total Energy vs. $M_{si}$ for different orientations of the modified region, for VW (open symbols) and TW (closed symbols) domain walls in the 500 nm wide, 10 nm thick wire.

From the observations presented in Section C we have identified a wide range of behavior for defects of different orientations. Taking these results together with our previous study, we can conclude that the overall trend is an increase in depinning field with dose for each defect angle together with a transformation of the pinned wall from transverse domain wall to vortex walls at higher doses. In Table I however, we record that more complicated behaviour is observed for linear defects irradiated at intermediate doses and we proceed to discuss this in detail.

**TABLE I.** $H_{dep}$ as a function of irradiation doses for pinning sites with different orientations of head-to head DW. In the initial transverse DW structure the magnetization points down (↓) as shown in (a). DW structure at the pinning site is also shown. CVW: clockwise vortex and CCVW: counterclockwise vortex. The table shows results from twelve measurements performed on each wire. Each measurement starts from the state shown in (a) and then follows the propagation and pinning (b), and depinning (c).

| Angle (deg) | 30 | 45 | 67.5 | 112.5 |
|---|---|---|---|---|
| **Dose ($10^{15}$ ions/cm$^2$)** | $H_{dep}$ (Oe) | $H_{dep}$ (Oe) | $H_{dep}$ (Oe) | $H_{dep}$ (Oe) |
| 4 | 9±2; TW↓ (100%) | 10±2; TW↑ (42%)<br>13±1; TW↓ (58%) | 7±2; TW↑ (8%)<br>9±1; TW↓ (92%) | 8±2; TW↑ (100%) |
| 8 | 26±5; CVW (100%) | 39±2; TW↑ (25%)<br>44±2; TW↓ (75%) | 25±2; TW↑ (17%)<br>30±2; TW↓ (83%) | 28±3; TW↓ (42%)<br>30±4; TW↑ (58%) |
| 12 | 47±1; CVW (100%) | 60±3; TW↓ (17%)<br>57±1; CVW (83%) | 42±3; TW↓ (83%)<br>40±5; CVW (17%) | 31±6; TW↓ (17%)<br>40±2; TW↑ (83%) |
| 16 | 66±1; CVW (100%) | 64±1; CVW (100%) | 50±3; CVW (100%) | 54±4; TW ↑ (17%)<br>50±8; CCVW (83%) |

We start by looking at an example of wall type changes as a function of angle for the lowest dose. From Table I a polarizing effect is seen on the initial transverse down wall (Fig. 8(a)) at the two extreme angles. In the case of the 30 degree irradiated defect, the leading edge of the wall exhibits a similar alignment with the linear defect and after propagation to the pinning site the wall remains as a transverse down type (Fig. 8(b)) . By contrast for the 112.5 degree defect, where the leading edge of the down wall is now poorly aligned to the defect (having an almost orthogonal alignment to the defect) the result is that the wall transforms to an up wall in order that leading edge better matches the defect (Fig. 8(c)). Between the two extrema, for intermediate angles a mixture of up and down walls is seen. However we need to differentiate between the up/down transverse walls and their position relative to the defect as shown in Figs. 8(b-e). In terms of the field direction they can be considered as pinned by their leading (Figs. 8(b) and (c)) or trailing (Figs. 8(d) and 8(e)) edges, i.e the up and down walls are different for line defects with orientation angles above and below 90 degrees. From Table I in all cases where we observe a mix of transverse walls at a particular dose and angle, we see a small but consistent difference between the depinning field for the up and down walls. What we note is that the depinning field is higher where the leading edge of the wall is pinned at the defect (Figs. 8(b) and 8(c)) whereas it is lower for the trailing edge pinned at the defect (Figs. 8(d) and 8(e)). Simplistically this makes sense in that the walls with the trailing edge are pulled from the defect whereas those pinned by the leading edge need to be pushed through the defect. Furthermore the chirality of the vortex domain wall now can be seen to be defined for angles below (CVW) and above 90 degrees (CCVW) as was observed in Fig. 6. This is also illustrated in Figs. 8(f) and 8(g) and is due to the alignment of the middle wall section with the defect line for a given chirality.

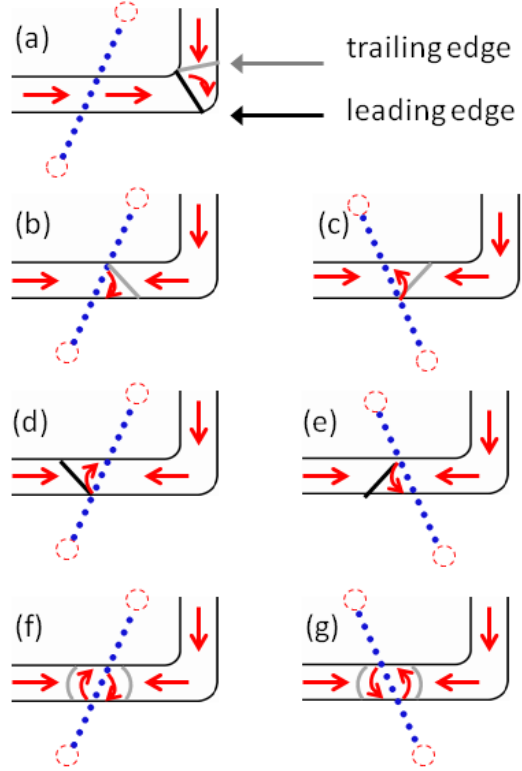

FIG 8. (color online). Schematics of the differing wall geometry (a) Initial transverse down wall. TWs with leading edge pinned to the defect: (b) down and (c) up. TWs with the trailing edge pinned to the defect: (d) up and (e) down. (f) and (g) VWs.

Another interesting observation is that the depinning field from the 45 degree line appears to be generally higher than for the other angles. Consideration of the total wall energy in fig, 7 may provide an explanation for such an effect. For walls in nanowires with no defect, the natural relaxed structure of both transverse and vortex type DWs favours a 45 degree orientation for the respective 90 degree (in the TDW) and 180 degree (in the VDW) Néel wall components that compose the total DW packet. Therefore with an irradiated defect positioned at this angle the individual component walls are well matched to the defects. However in the case with the defects at angles other than 45 degrees the component walls must distort to align to the defect suggesting that energy will be significantly higher than for defects at 45 degrees. Figure 7 shows that the total energy of the transverse and vortex DWs is calculated to be lowest for all of the 45 degree defects. Thus, the 45 degree structure may be regarded as the deepest potential well and requiring a greater field to extract the DW from this potential.

Finally we discuss the appearance of vortex walls for higher irradiation doses as recorded in Table I. With increasing irradiation dose, figure 7 showed that the energy of TWs decreases more rapidly than for VWs, and for irradiation causing $M_{si} \leq 0.1 M_s$, VW and TW energies become equivalent. In figure 9 we show micromagnetic simulations of head-to-head TW and VWs at the irradiated defect. When $M_{si} = 0.9 M_s$ the TW, figure 9(a), and the VW, figure 9(b), exhibit their commonly observed structures. Corresponding to irradiation at a high dose, when $M_{si} = 0.3 M_s$, figure 9(c) shows the simulated structure of the TW becomes considerably modified compared to the usual TW with the structure becoming more like the VW of figure 9(d) but with a significantly reduced transverse component of magnetization. For the modified TW in figure 9(c) to become the VW of figure 9(d) a vortex requires to be nucleated. This is a process that we would expect to occur easily as a result of thermal activation - the vortex being nucleated at the top edge of the wire and immediately travelling along the 180 degree Néel wall line to occupy a central position. Experimentally, the appearance of the vortex wall for higher angle defects ($\geq 45$ degrees) is observed for irradiation doses $\geq 12 \times 10^{15}$ ions/cm$^2$ which corresponds to $M_{si} < 0.3 M_s$ as shown in Fig. 4 (d). The appearance of the vortex wall at much lower dose for the 30 degree defect compared to the others is attributed to the fact that an extension of the wall is needed for the transverse wall to align to the defect. Thus the transformation from the metastable transverse wall to a presumably ground state vortex structure is more likely for lower doses.

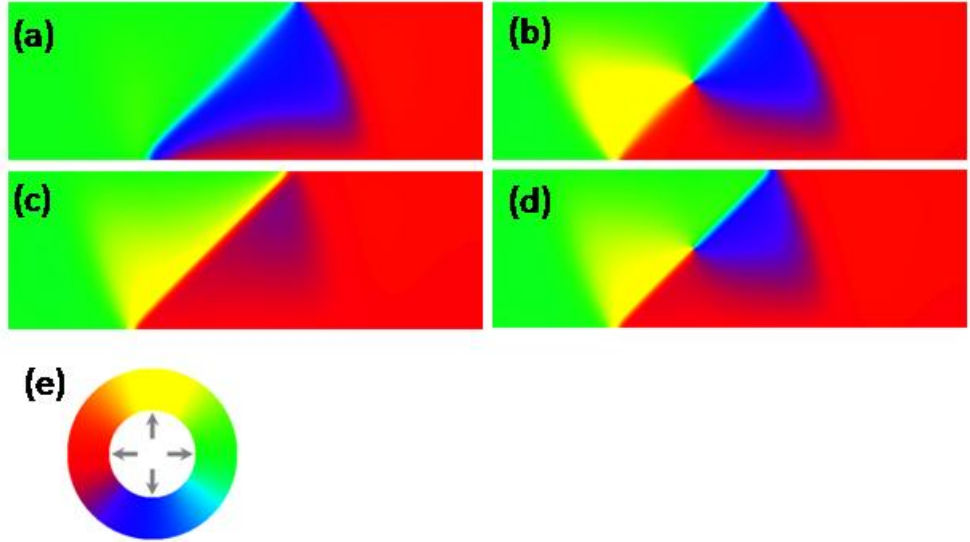

FIG 9. (color online). Calculated color images from simulations of the wires with local property modification. (a) shows a TW and (b) a VW with $M_{si} = 0.9M_s$, (c) shows the modified TW and (d) the VW with $M_{si} = 0.3M_s$. (e). Color wheel.

## IV. CONCLUSIONS

In summary, our results demonstrate that the structure of DWs in these Permalloy multi-layer nanowires has been "engineered" by local property modification via the linear defect region. The angle of the linear defect and/or irradiation dose can serve to stabilizse DW structures that are quite different to the equilibrium structure that one would find in an unmodified nanowire. TW are favored at lower doses and larger defect angles whereas VWs are predominant at the highest doses and lower angles. By changing the angle of the irradiated line it is possible to vary the DW packet width, pinning strength and vortex chirality. Additionally the observed transverse walls were noted to have different pinning behavior depending on whether there were pinned with their leading or trailing edge at the defect. Quantification of the effects of the $Ga^+$ ion irradiation were made possible using the methods of Lorentz microscopy, in particular using DPC imaging we show that it is possible to detect changes in magnetic induction for samples with locally reduced saturation magnetization on a lengthscale of Néel domain wall widths (a few tens of nanomteres).

ACKNOWLEDGEMENTS

We would like to thank W. A. Smith and C. R. How for their expert technical assistance. This research is supported by the Engineering and Physical Sciences Research Council, EPRSC (grant numbers EP/I013520 and EP/I011668/1), the Scottish Universities Physics Alliance and the University of Glasgow.